\documentclass[]{ceurart}

\usepackage{mathptmx}
\usepackage{soul}\setuldepth{article}
\usepackage{enumitem}
\usepackage{url}

\def\hb{\hbox to 11.5 cm{}}

\usepackage{graphicx}
\usepackage{comment}
\usepackage{url} 

\usepackage{tabulary} 
\usepackage{booktabs} 

\usepackage{listings} 
\usepackage{tabularx} 
\usepackage{nameref} 
\usepackage{graphicx}
\usepackage[flushleft]{threeparttable} 
\usepackage{svg} 
\usepackage[strict]{changepage}
\usepackage{framed} 
\usepackage{xspace} 
\usepackage{subcaption}

\usepackage[numbers]{natbib}

\newcommand{\lkgb}{LLM-KG-Bench\xspace}
\newcommand{\sparqlselquery}{\emph{SPARQL SELECT} query\xspace}
\newcommand{\sparqlselqueries}{\emph{SPARQL SELECT} queries\xspace}
\newtheorem{researchQuestion}{RQ} 

\definecolor{formalshade}{rgb}{0.95,0.95,1}
\definecolor{darkblue}{rgb}{0.0, 0.0, 0.55}

\newcounter{promptnumber}
\newenvironment{prompt}{%
 \refstepcounter{promptnumber}%
  \MakeFramed{\advance\hsize-\width\FrameRestore}%
  \noindent\hspace{-4.55pt}
  \begin{adjustwidth}{}{7pt}%
  \vspace{2pt}\vspace{2pt}%
   \textbf{Prompt \thepromptnumber:}~
}
{%
  \vspace{2pt}\end{adjustwidth}\endMakeFramed%
}

\usepackage{hyperref}
\usepackage{cleveref}
\Crefname{promptnumber}{Prompt}{Prompts}%
\crefname{promptnumber}{prompt}{prompts}%

\begin{document}

\copyrightyear{2024}
\copyrightclause{Copyright for this paper by its authors.
  Use permitted under Creative Commons License Attribution 4.0
  International (CC BY 4.0).}

\conference{NLP4KGC: 3rd International Workshop on Natural Language Processing for Knowledge Graph Creation, \mbox{September}~17, 2024, Amsterdam, Netherlands}

\title{Assessing SPARQL Capabilities of Large Language Models}

\author[A,B]{Lars-Peter Meyer} [orcid=0000-0001-5260-5181, email=lpmeyer@infai.org]
\cormark[1]
\author[A,B]{Johannes Frey} [orcid=0000-0003-3127-0815, email=frey@informatik.uni-leipzig.de]
\author[A]{Felix Brei} [orcid=0009-0008-5245-6655, email=brei@infai.org]
\author[A]{Natanael Arndt} [orcid=0000-0002-8130-8677]

\address[A]{Institute for Applied Informatics at Leipzig University, Goerdelerring 9, 04109 Leipzig, Germany}
\address[B]{Leipzig University, Germany}
\cortext[1]{Corresponding author.}

\begin{abstract}
The integration of Large Language Models (LLMs) with Knowledge Graphs (KGs) offers significant synergistic potential for knowledge-driven applications.
One possible integration is the interpretation and generation of formal languages, such as those used in the Semantic Web, with SPARQL being a core technology for accessing KGs.
In this paper, we focus on measuring out-of-the box capabilities of LLMs to work with SPARQL and more specifically with SPARQL SELECT queries applying a quantitative approach.

We implemented various benchmarking tasks in the LLM-KG-Bench framework for automated execution and evaluation with several LLMs.
The tasks assess capabilities along the dimensions of syntax, semantic read, semantic create, and the role of knowledge graph prompt inclusion.

With this new benchmarking tasks, we evaluated a selection of GPT, Gemini, and Claude models.
Our findings indicate that working with SPARQL SELECT queries is still challenging for LLMs and heavily depends on the specific LLM as well as the complexity of the task.
While fixing basic syntax errors seems to pose no problems for the best of the current LLMs evaluated, creating semantically correct SPARQL SELECT queries is difficult in several cases.
\end{abstract}

\begin{keywords}
    LLM benchmarking \sep
    SPARQL \sep
    LLMs for Knowledge Graphs \sep
    RDF \sep
    LLM \sep
    Knowledge Graph
\end{keywords}

\maketitle

\begin{picture}(0,0)
  \put(\dimexpr\paperwidth-2in\relax,0.4in){%
    \makebox(0,0)[r]{\rotatebox{90}{\LARGE peer reviewed and published in \href{https://sites.google.com/view/3rdnlp4kgc}{NLP4KGc} @ \href{https://2024-eu.semantics.cc/}{SEMANTiCS 2024} workshop \href{https://ceur-ws.org/Vol-3874/paper3.pdf}{proceedings}}}%
  }
\end{picture}
%
%
\section{Introduction}

The combination of Large Language Models (LLMs) and Knowledge Graphs (KGs) is still gaining more traction as the rapidly growing number of research articles\footnote{list of articles combining KG+LLM: \url{https://github.com/zjukg/KG-LLM-Papers}} and the inclusion of LLMs in several KG related conference calls show.
This includes different combinations \cite{PanUnifyingLLMsAndKGs} of LLMs and KGs, especially LLMs supported by KGs and LLMs supporting the work with KGs.
An important interface for both combinations is the SPARQL query language as the structured retrieval and update interface for RDF KGs \cite{Hogan2020KnowledgeGraphs}.

With the fast evolving list of LLMs available, the need for automated benchmarks increases.
LLM performances is evaluated by BigBench \cite{srivastava2023imitation}, on the Open LLM Leaderboard \cite{Open-LLM-Leaderboard-Report-2023}, and the Chatbot-Arena \cite{chiang2024chatbot}, but these are very general and do not focus on KGs or SPARQL.
In existing work, KG related benchmarking is done specifically on the Turtle format \cite{Frey2023Turtle} or often as discussion and comparison of very specific solutions.
LLM benchmarking related to SPARQL was done in a small scale by Meyer et al. \cite{Meyer2023ExperimentsWithChatGPT} and by Kovriguina et al. \cite{Kovriguina2023SPARQLGENOP}.

In this paper, a set of automated benchmarking tasks is presented, to assess the basic capabilities of LLMs to deal with \sparqlselqueries without special architectures or fine tuning.
We reuse existing work on Knowledge Graph Question Answering (KGQA) datasets, taking lists of question SPARQL pairs for given KGs.
The benchmarking tasks are implemented in the \lkgb 
framework\footnote{repository: \url{https://github.com/AKSW/LLM-KG-Bench}}\cite{meyer2023developing}.
The tasks are executed on selected LLMs and the results are presented and discussed.

\subsection{Research Questions}

In this paper, we consider the following research questions.

\begin{researchQuestion} \label{rq:SparqlSyntaxReadWrite}
    Can LLMs follow the syntactic rules of \sparqlselqueries?
\end{researchQuestion}
\begin{researchQuestion} \label{rq:SemanticRead}
    Can LLMs parse the semantics of \sparqlselqueries and act accordingly?
\end{researchQuestion}
\begin{researchQuestion} \label{rq:SparqlWrite}
    Can LLMs write semantically correct \sparqlselqueries for a given question and KG, i.e. a query that yields the expected answer?
\end{researchQuestion}
\begin{researchQuestion} \label{rq:KgImpact}
    Does the KG presentation have an influence on the ability to write \sparqlselqueries?
\end{researchQuestion}

\begin{figure}
  \begin{center}
  \includegraphics{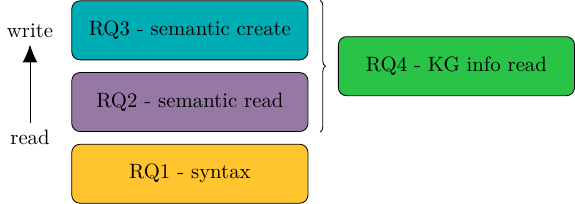}
  \end{center}
  \caption{An overview on the relation of the research questions.}
  \label{fix:rq}
\end{figure}

The research questions depend on each other and build on the previous questions as depicted in \cref{fix:rq}.
RQ1 focusses on the syntax, while RQ2 and RQ3 aim at semantics. RQ2 covers the interpretation and RQ3 the creation of said queries.
RQ4 examines the influence of the way knowledge is presented on the results of RQ2 and RQ3.

\subsection{Contribution}

The contribution of the paper is both an evaluation of current LLMs as well as the extended evaluation framework itself.
Our evaluation and framework is comprised of tasks that are executed in various configurations.
In section \ref{sec:taskTypesDescriptions} we introduce the tasks.
The main task types are \textit{SPARQL Syntax Fixing (SSF)}, \textit{Text to SPARQL translation (T2S)}, \textit{SPARQL to Answer (S2A)} and \textit{Text to Answer (T2A)}. These tasks evaluate relevant aspects to cover the dimensions along which the capabilities (see \cref{sec:taskAspectsDescription}) of the models are assessed according to our research questions:
\begin{itemize}
    \item Task type SSF focuses on working with the syntax of \sparqlselqueries, giving answers for RQ1.
    \item Task type S2A is asking to work according to semantics of \sparqlselqueries, helping with RQ2.
    \item In task type T2S the LLM should create syntactically and semantically correct \sparqlselqueries, giving us answers to RQ1, RQ3 and RQ4.
    \item For answering task type T2A the LLM needs to get information from the KG, thus the results are relevant for RQ4.
\end{itemize}
The purpose of RQ4 is to understand the influence of the knowledge graph's syntax (Turtle, JSON-LD), the representation of the data (numerical identifiers, human readable identifiers), and the knowledge graph prompt inclusion (schema, size of the sub graph, and selection) in the answers to RQ2 and RQ3.
The variants of execution with Turtle and JSON-LD as well as the selection of datasets (KG info type) and their subsets pay in on RQ4.

The remainder of the paper is structured as follows:
In \cref{sec:relwork} an overview on the related work is given and the paper is located with its contribution to the bigger picture.
\Cref{sec:expsetup} presents the kind of tasks that were executed as part of our benchmark and which datasets were used.
In \cref{sec:results} we do an extensive evaluation of the obtained results and the conclusions that can be drawn.
Section \ref{sec:conclusion} summarizes the findings and gives an overview on the directions that our research will head next.

\section{Related Work}
\label{sec:relwork}

Rangel et al.~\cite{Reyes2024SPARQLGA} propose a methodology for fine-tuning OpenLLaMA to generate SPARQL queries for question answering over life science knowledge graphs using data augmentation techniques, such as providing meaningful variable names and inline comments, improving the performance of the model in generating accurate SPARQL queries.
Bustamante and Takeda \cite{Bustamante2024SPARQLGW} aim at improving the creation of SPARQL queries based on natural language questions. The authors use a GPT model to identify which property of the Text2SPARQL task is the hardest to solve, in order to select appropriate countermeasures.
Avila et al.~\cite{Avila2024T2S} evaluate the ability to answer natural language questions with ChatGPT on KGs. With the Auto-KGQAGPT approach the authors further investigate the ability to translate natural language question to SPARQL queries based on a prompt inclusion of KG fragments.

Li et al.~\cite{Li2023FlexKBQAAF} address the challenge of declined quality of results in real-world scenarios where high-quality annotated data is insufficient.
To target this challenge the authors present the FlexKBQA framework which employs templated SPARQL queries that are translated into natural language questions using LLMs to generate synthetic training data. This synthetic data allows to fine-tune a lightweight model for SPARQL generation along with further self-guided training on real queries to address a distribution shift between synthetic and real queries.

Diallo et al.~\cite{Diallo2023ACE} give a comprehensive overview and performs a comparison of pre-trained LMs (PLMs), non-pre-trained LMs (NPLMs), and LLMs, testing various fine-tuning methods using LLMs. The error analysis of the models results in the finding, that the primary source of errors are incorrect URIs in SPARQL queries e.g. due to hallucination.
Hirigoyen et al.~\cite{hirigoyen-etal-2022-copy} have found though that the hallucination can be prevented by using a copy mechanism.

The SPARQLGEN approach by Kovriguina et al.~\cite{Kovriguina2023SPARQLGENOP} is a one-shot generative approach to generate SPARQL queries using LLMs. It includes the relevant context in a single prompt, i.e. the question, an RDF sub graph with the relevant information, and an example of a SPARQL query and a question.
Pliukhin et al.~\cite{Pliukhin2023ImprovingSE} present an approach for SPARQL query generation on the scholarly knowledge graph ORKG. Its setup is similar to SPARQLGEN but utilizes an advanced sub graph extraction.
Zahera et al.~\cite{Zahera2024GeneratingSPARQLNatural} went on with leveraging chain of thougth prompting and utilizes the GERBIL QA system\footnote{projekt page: \url{https://gerbil-qa.aksw.org/gerbil/}}\cite{Usbeck2019Benchmarkingquestionanswering}.
Lehmann et al. \cite{Lehmann2023CNL} suggest the usage of Controlled Natural Language as a human interface since it is closer to natural language. The Controlled Natural Language can then be unambiguously translated into a formal language such as SPARQL. They come to the conclusion that this approach substantially reduces the training data requirements.

Several datasets of complex questions over knowledge graphs are available.
Diefenbach et al.~\cite{DiefenbachTSM17} present two datasets for training and benchmarking question answering systems using Wikidata. One is a translation of the SimpleQuestions dataset (cf.~Bordes et al.~\cite{Bordes2015LargescaleSQ}), the other is based on logs and user feedback.
LC-QuAD 2.0 by Dubey et al.~\cite{dubey2017lcQuad2} is an advancement of the Large-Scale Complex Question Answering Dataset (LC-QuAD) \cite{Trivedi2017LCQuADAC}. It is compatible with both, the Wikidata and the DBpedia knowledge graphs. It contains “30,000 questions, their paraphrases and their corresponding SPARQL queries”.

Banerjee et al. \cite{Banerjee_2022} evaluate different approaches to incorporate LLMs into a common KG workflow. They use language models that can be fine tuned on consumer hardware today, while we focus on very large models that will be used interactively and as is, without any fine-tuning.

Frey et al. \cite{Frey2024AssessingEvolutionLLM} show the long term value of automated benchmarking with \lkgb. 
Several version iterations of LLMs were compared with focus on their RDF Turtle language capabilities and the answers were conserved in a time capsule for future (re)evaluation. 
Hofer et al.~\cite{hofer2022towards} showed in an LLM-driven RML mapping (in Turtle format) generation experiment  in alignment with \cite{Frey2024AssessingEvolutionLLM,Frey2023Turtle}, that syntactical errors occurred when generating Turtle files, but most LLMs were capable to repair them. 
Given that SPARQL is based on Turtle, we designed our experiments as multi turn conversations with error feedback loops.

\section{Experiment Setup}
\label{sec:expsetup}

For automated assessment of the capabilities of LLMs to work with \emph{SPARQL SELECT} statements we extended the \lkgb framework. In the following subsections we outline evaluation methods, the different task types, KGs used for testing and specific tasks.

 nomenclature that we used:

\begin{itemize}
    \item A \textit{task type} is a class of tasks that an LLM needs to solve, like SSF or T2S.
    \item A \textit{task} is one specific instantiation of a task type, like T2A on LC-QuAD using JSON-LD syntax for the knowledge graph.
    \item \textit{Task entry} refers to one single data point that is used during a task, for example the third question in T2A for LC-QuAD.
    All tasks presented here have a list of five task entries which they choose from evenly distributed.    
    \item One \textit{execution} of a task entry consists of all steps that are necessary to reach completion of this task entry. For example, the execution of one task entry in SSF entails the complete feedback dialog and ends with the final answer of the LLM and its evaluation.
\end{itemize}

\subsection{Experiment Type Selection}

\begin{figure}
    \centering
    \includegraphics{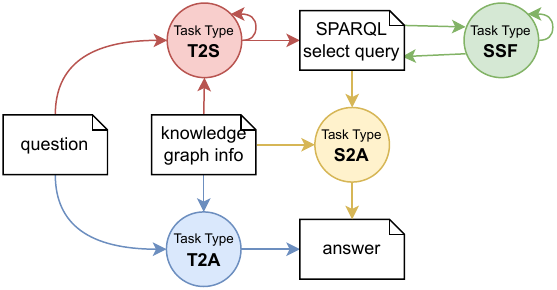}
    \caption{Overview on the task types and their input and output}
    \label{fig:sparqlTaskTypes}
\end{figure}

\emph{SPARQL SELECT} queries are connected to a related textual question, a knowledge graph the query is designed for, and an answer.
For working with SPARQL SELECT queries one usually has to deal with a combination of these concepts.
This leads us to the four different task types with inputs and outputs as shown in \cref{fig:sparqlTaskTypes}.

\subsection{Detailed Explanation of Task Types}
\label{sec:taskTypesDescriptions}

\subsubsection{Task Type ``SPARQL Syntax Fixing'' (SSF)}
The task SPARQL Syntax Fixing asks to fix a syntax error in a given SPARQL query.
To be more precise, the LLM is given a syntactically malformed SPARQL query together with the resulting rdflib\footnote{\url{https://github.com/RDFLib/rdflib}} parsing error message.
The LLM is asked to answer with just the fixed SPARQL query.

The following prompt is used with variables(\verb|${sparql}| and \verb|${errorMessage}|) filled in according to the specific task:
\begin{prompt}
    Please correct a syntax error in the following SPARQL query for Wikidata. Assume common prefixes like wd or wdt to be defined.
    To support automated parsing, please answer with just a markdown fenced code block (start and end with ```) containing the sparql query, no other text.\\
    \\
    Example for Answer format:\\
    \verb|```|sparql\\
    SELECT ...\\
    \verb|```|\\
    \\
    SPARQL:\verb|${sparql}|\\
    \\
    Error message: \verb|${errorMessage}|\\
\end{prompt}

The answer presented is evaluated according to remaining syntax errors and having the expected result set when executed as described in \cref{sec:SparqlAnswerEvaluation}.
The LLM is given the opportunity to correct the answer in a feedback dialog session with up to three tries.

\subsubsection{Task Type ``Text to SPARQL'' (T2S)}
The Text-to-SPARQL task type asks to create a \emph{SPARQL SELECT} query for a KG and a textual question.
Many KGQA approaches build upon this capability.

The following prompt is used with several variables(\verb|${kgName}|, \verb|${commonPrefixes}|, \verb|${question}| and \verb|${KgInfo}|) filled in according to the specific task:
\begin{prompt}
    Please generate a SPARQL query for \verb|${kgName}| and the given question. Assume common prefixes \verb|${commonPrefixes}| to be defined.
    To support automated parsing, please answer with just a markdown fenced code block (start and end with ```) containing the sparql query, no other text.\\
    \\
    Example for Answer format:\\
    \verb|```|sparql\\
    SELECT ...\\
    \verb|```|\\
    \\
    Question:\verb|${question}|\\
    \\
    \verb|${KgInfo}|\\
\end{prompt}

The given SPARQL SELECT query is evaluated on being syntactically correct and having the same result set as expected as described in \cref{sec:SparqlAnswerEvaluation}.
The LLM is given the opportunity to correct the answer in a feedback dialog session with up to three tries.

\subsubsection{Task Type ``SPARQL to Answer'' (S2A)}
The SPARQL-to-Answer task asks to interpret a given \emph{SPARQL SELECT} query on a given KG and answer with the binding values.
The LLM is asked to give one value per line and the F1 measure is used to score the result, as described in more detail in \cref{sec:TextAnswerEvaluation}.

\subsubsection{Task Type ``Text to Answer'' (T2A)}
This task type was added as a variation of the previous ``SPARQL to Answer`` task type for comparison.
The task is to answer a given natural language textual question on a given KG and answer with the binding values.
Expected answer structure and evaluation is the same as described above for task type S2A.

\subsection{Evaluation Method}
In this paper we are applying a quantitative approach.
To support automated evaluation within in the \lkgb framework we need to define the following evaluation methods for \sparqlselqueries as generated by task types SSF and T2S, and for answers given in task types S2A and T2A.

\subsubsection{Evaluation of SPARQL SELECT Queries (SSF and T2S)}
\label{sec:SparqlAnswerEvaluation}
\sparqlselqueries are evaluated regarding having correct syntax and yielding the expected result when executed on the KG.
To be more precise, all entries of the result bindings are put into a set (duplicates removed) and are then compared to the expected result set.
In case of syntax errors or an empty result set the LLM is given the opportunity to correct the \sparqlselquery in a feedback dialog session.
Therefore a feedback message is generated and sent back to the LLM together with the previous dialog.
This feedback dialog session is continued until up to three answers (one initial and up to two additional tries for corrections) are collected or  at least one \sparqlselquery yields a result.

In case of syntax errors the following prompt template was used with values filled in for \verb|${error}| and \verb|${sparql}|:
\begin{prompt}
    Please try to correct your answer. Your SPARQL query has syntax errors: \verb|${error}|\\
    \\
    SPARQL given:\\
    \verb|```|sparql\\
    \verb|${sparql}|\\
    \verb|```|
\end{prompt}

In case of an empty result the following prompt was used:
\begin{prompt}
    Maybe you want to think again about your answer. Your SPARQL query returns an empty result when executed.
\end{prompt}

The following scores are generated:
\begin{description}
    \item[answerParse] $\in[0,1]$:
        1 if the \sparqlselquery given can be parsed with rdflib, 0 otherwise
    \item[\{f1measure\textbar precision\textbar recall\}] $\in[0..1]$: comparison of the result set given and expected.
    \item[sparqlIris\{f1measure\textbar precision\textbar recall\} ] $\in[0..1]$:
        comparison of the IRIs in the \sparqlselquery given and expected.
    \item[sparqlIriSuffix\{f1measure\textbar precision\textbar recall\} ] $\in[0..1]$:
        comparison of the last segments(the part after last \verb|#| or \verb|\|) of the IRIs in the \sparqlselquery given and expected.
    \item[combined] $= 0.2 * parse + 0.8 * f1measure$ $\in[0..1]$: combining syntax and semantic score, resulting e.g. in a score of $0.2$ when f1 measure on result set is $0$ but the \sparqlselquery parses.
\end{description}
The values above get prefixed with the specified iteration (\textbf{0\_\textellipsis}, \textbf{1\_\textellipsis}, \textbf{2\_\textellipsis}, \textbf{last\_\textellipsis}) of the feedback dialog session and aggregates (\textbf{mean\_\textellipsis}, \textbf{max\_\textellipsis}) across the whole feedback session are calculated, giving us score names like \verb|0_answerParse|, \verb|1_answerParse|, \dots as well as \verb|mean_f1measure| or \verb|max_precision| to name a few.

\subsubsection{Evaluation of Answers Given by Tasks S2A and T2A}
\label{sec:TextAnswerEvaluation}
In the task types S2A and T2A the LLM is asked to give the answer from the KG similar to binding values, one value per line.
For comparing the given answer lines with the expected answer lines, precision, recall and f1 measure are calculated.
As the LLMs answer does not always follow the expected structure perfectly we implemented a couple of scores for strict and more flexible answer parsing as well:
\begin{description}
    \item[exact (f1, precision, recall) ] $\in[0..1]$: comparison on the exact lines given with expected answer entries
    \item[trimmed (trimF1, trimPrecision, trimRecall) ] $\in[0..1]$: comparison after leading and trailing whitespace is removed
    \item[fixed format (fixedF1, fixedPrecision, fixedRecall) ] $\in[0..1]$: comparison after fixing simple format errors like http instead of https, prefix \verb|0:| (often given for JSON-LD) instead of \verb|:| and removal of brackets (\verb|<| and \verb|>|) and quotation marks (\verb|'| and \verb|"|)
    \item[relaxed evaluation (relaxedF1, relaxedPrecision, relaxedRecall) ] $\in[0..1]$: comparison with format fixed plus ignore case, removing default prefix \verb|:| (makes instances similar to labels for semantic IRIs). In case a count is expected a list with the right length is also accepted as correct.
    \item[combinedF1] \(= \frac{f1 + trimF1 + fixedF1 + relaxedF1}{4} \in[0..1]\): mean of the f1 scores above.

\end{description}

\subsection{Task Aspects Covered by the Task Types}
\label{sec:taskAspectsDescription}
The 4 task types described above in \cref{sec:taskTypesDescriptions} are covering a list of task aspects and are relevant for the research questions. They are mapped on task types in \cref{tab:taskTypes} and described in the following list:
\begin{description}

    \item[read \sparqlselquery syntax:]
        To work with a \sparqlselquery the syntax must be digested.

        Part of task types SSF and S2A, related to RQ 1.

    \item[create correct \sparqlselquery syntax:]
        Create or write syntactically correct \sparqlselquery syntax.

        Part of task types SFF and T2S, related to RQ 1.

    \item[read semantics of a \sparqlselquery:]
        Get the semantic meaning of a \sparqlselquery and act accordingly.

        Part of task type S2A, related to RQ 2.

    \item[create a semantically correct \sparqlselquery:]
        create a \sparqlselquery not only syntactically correct but with an appropriate semantic which yields the expected bindings.

        Part of task type T2S, related to RQ 3.

    \item[KG info read]:
        When dealing with a knowledge graph some kind of information on the KG structure and KG content is needed.
        The information on the structure could be anything from whole KG over schema to just a list of relevant or all entities and properties.

        Part of task types T2S, T2A and S2A, related to RQ 4.

\end{description}

\begin{table}
    \caption{
        Mapping of the task types to the different task aspects covered.
    }
    \begin{tabulary}{\textwidth}{LC@{\hskip 5pt}C@{\hskip 5pt}C@{\hskip 5pt}C@{\hskip 5pt}C}
        \toprule
    task type              & \multicolumn{5}{l}{task aspects}     \\
        \cmidrule(l){2-6}
                           & syntax read & syntax create & semantic read & semantic create & KG info read \\
        \midrule
SPARQL Syntax Fixing (SSF) & x                 & x                   & -                   & -                     & -                  \\
SPARQL to Answer (S2A)     & x                 & -                   & x                   & -                     & x                  \\
Text to SPARQL (T2S)       & -                 & x                   & -                   & x                     & x                  \\
Text to Answer (T2A)       & -                 & -                   & -                   & -                     & x                  \\
        \bottomrule
    \end{tabulary}
    \label{tab:taskTypes}
\end{table}

\subsection{Benchmark Datasets and KGs Used for Task Implementations}
There are several benchmark datasets available which contain pairs of \sparqlselqueries and textual natural language questions for specific knowledge graphs.
We selected and implemented benchmark tasks for a couple of current datasets for smaller and bigger knowledge graphs.
Only English textual questions were used as we focus on \emph{SPARQL} here and not language capabilities.
A total of five tuples, each consisting of a question and corresponding SPARQL query, was manually selected from each dataset. This allows to rerun the tasks more often to reduce the random noise in the results.

\begin{description}
    \item[Organizational dataset and KG]
        The smallest KG used is an organizational KG \cite{Meyer2023ExperimentsWithChatGPT}.
        We use it here together with a corresponding dataset\footnote{Repository: \url{https://github.com/AKSW/LMs4Text2SPARQL/tree/main/datasets}} of question and SPARQL pairs created by Brei et al. \cite{Brei2024LeveragingSmallLanguage}.
    \item[Organizational Numeric]
        We created an additional variant of the organizational dataset and KG with numerical IRIs (first 3 digits of hash) and same questions.
    \item[CoyPu-Mini dataset and KG]
        Brei et al. \cite{Brei2024LeveragingSmallLanguage} published as well another dataset based on a small subset of the CoyPu KG\footnote{Project page: \url{https://coypu.org/}}.
        This sub graph is small enough to fit into context size of LLMs evaluated here.
        We added lists of relevant IRIs and schema information.
    \item[Beastiary dataset and KG]
        The Beastiary dataset and KG\footnote{Repository: \url{https://github.com/danrd/sparqlgen}} was presented by Kovriguina et al. \cite{Kovriguina2023SPARQLGENOP}.
        It offers for each question a relevant sub graph and a list of IRIs.
        We derived relevant schema information from these IRI lists.
    \item[LC-QuAD 2.0 and Wikidata]
        The well known \emph{LC-QuAD 2.0} \footnote{Dataset: \url{https://huggingface.co/datasets/lc_quad}} \cite{dubey2017lcQuad2} dataset offers a long list of question and SPARQL pairs for the Wikidata SPARQL endpoint.
        We focused on the \emph{paraphrased questions} of the test dataset and manually checked them before selecting 5 out of the first 20.
        Main reason for rejecting question was a missing or ambiguous paraphrased question.
        We computed lists of relevant IRIs for each question based on the reference SPARQL query.
        As some questions do not specify whether an IRI or label or both is expected, we configured the evaluation to consider all variations as correct there.
    \item[SPARQL SELECT Query Syntax Errors]
        We took one question and SPARQL pair from LC-QuAD and derived 5 tests from it by inserting different kinds of syntax errors, one error per test.
\end{description}

The list of benchmark tasks created based on this resources in the \emph{\lkgb} framework is given in \cref{tab:taskList}.

\begin{table}
    \centering
    \caption{Overview on the different tasks implemented within \lkgb framework}
    \begin{tabulary}{\textwidth}{LLCCC}
        \toprule
        task type~~~ & dataset subset         & task                & KG Info type      & KG info format    \\
        \midrule
        SSF       & Syntax-Errors          & SSF-LC-QuAD          & not needed here   & not needed here   \\
        \cmidrule(r){1-2}
        S2A       & Organizational         & S2A-Orga            & full KG           & Turtle or JSON-LD \\
        \cmidrule(r){1-2}
        T2S       & LC-QuAD 2.0            & T2S-LC-QuAD          & IRIs + Labels       & table             \\
        \cmidrule(r){2-2}
                  & Organizational         & T2S-Orga            & full KG           & Turtle            \\
        \cmidrule(r){2-2}
                  & Organizational-Numeric & T2S-OrgaNum         & full KG + IRIs + Labels & Turtle + table    \\
        \cmidrule(r){2-2}
                  & Beastiary              & T2S-Beast-Graph     & KG subset         & Turtle            \\
                  &                        & T2S-Beast-Schema    & schema            & Turtle            \\
                  &                        & T2S-Beast-Subschema & schema subset     & Turtle            \\
                  &                        & T2S-Beast-Iris      & IRIs              & list              \\
        \cmidrule(r){2-2}
                  & CoyPu-Mini             & T2S-CoyPu-Graph     & full KG           & Turtle or JSON-LD \\
                  &                        & T2S-Coypu-Schema    & schema            & Turtle or JSON-LD \\
                  &                        & T2S-Coypu-Iris      & IRIs              & list              \\
        \cmidrule(r){1-2}
        T2A       & Organizational         & T2A-Orga            & full KG           & Turtle or JSON-LD \\
        \bottomrule
    \end{tabulary}
    \label{tab:taskList}
\end{table}

\section{Results}
\label{sec:results}

We evaluated the LLMs from OpenAI, Anthropic and Google listed in \cref{tab:LLMs} with tasks listed in \cref{tab:taskList}.
Each combination of LLM and task shown here was executed 50 times, resulting in 10 executions per task entry and LLM.
All data generated is published at GitHub\footnote{link is given at the end of paper}. The evaluation can be reproduced by anyone by using the \verb|--reeval| parameter of the \lkgb framework.

\begin{table}
  \centering
  \caption{LLMs evaluated}
  \begin{tabular}{llll}
    \toprule
    company   & short name       & long name                & context size  \\
    \midrule
    Anthropic & Claude 2.1       & Claude-2.1               & 200k          \\
              & Claude 3 Haiku   & claude-3-haiku-20240307  & 200k          \\
              & Claude 3 Sonnet  & claude-3-sonnet-20240229 & 200k          \\
              & Claude 3 Opus    & claude-3-opus-20240229   & 200k          \\
    \cmidrule(r){1-1}
    Google    & Gemini 1.0       & gemini-1.0-pro           &               \\
              & Gemini 1.5       & gemini-1.5-pro-latest    &               \\
    \cmidrule(r){1-1}
    OpenAI    & GPT 3.5t 2024/01 & gpt-3.5-turbo-0125       & 16k           \\
              & GPT 4t 2023/11   & gpt-4-1106-preview       & 128k          \\
              & GPT 4t 2024/04   & gpt-4-turbo-2024-04-09   & 128k          \\
    \bottomrule
  \end{tabular}
  \label{tab:LLMs}
\end{table}

\begin{figure}[bt]
  \centering
  \begin{subfigure}{0.24\textwidth}
    \centering
    \includegraphics[width=\linewidth]{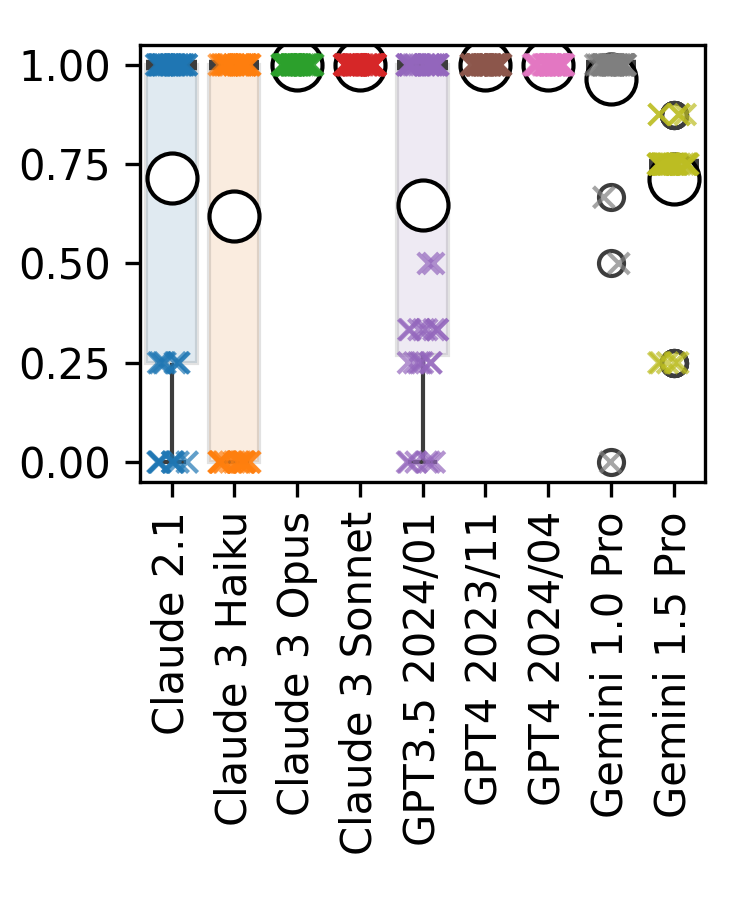}
    \caption{S2A JSON-LD}
    \label{fig:S2A-Orga-Jsonld}
  \end{subfigure}%
  \hfill
  \begin{subfigure}{0.24\textwidth}
    \centering
    \includegraphics[width=\linewidth]{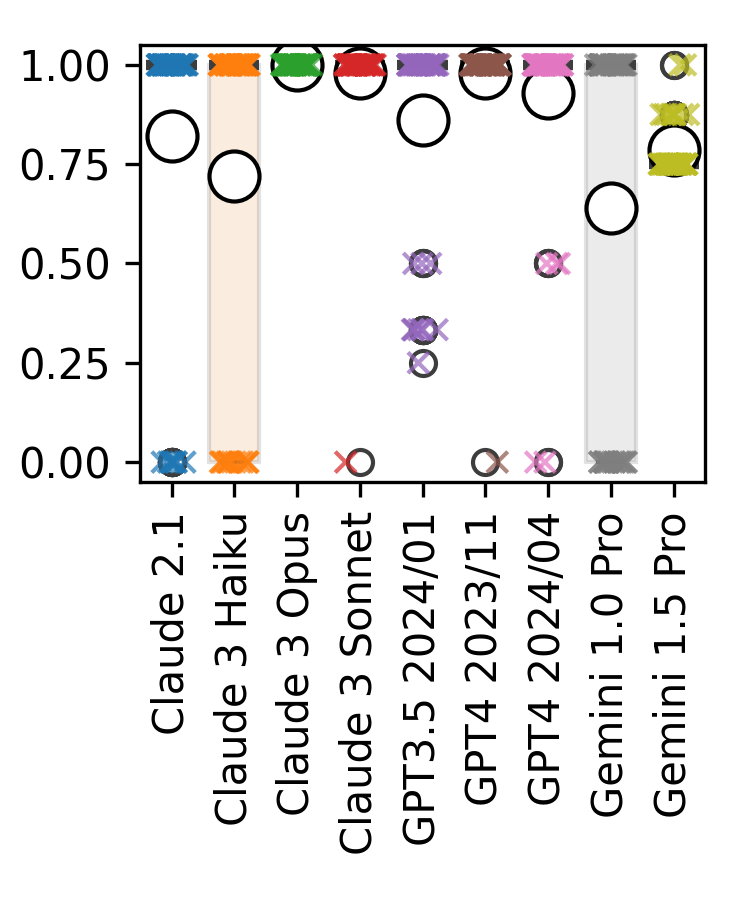}
    \caption{S2A Turtle}
    \label{fig:S2A-Orga-Turtle}
  \end{subfigure}
  \hfill
  \begin{subfigure}{0.24\textwidth}
    \centering
    \includegraphics[width=\linewidth]{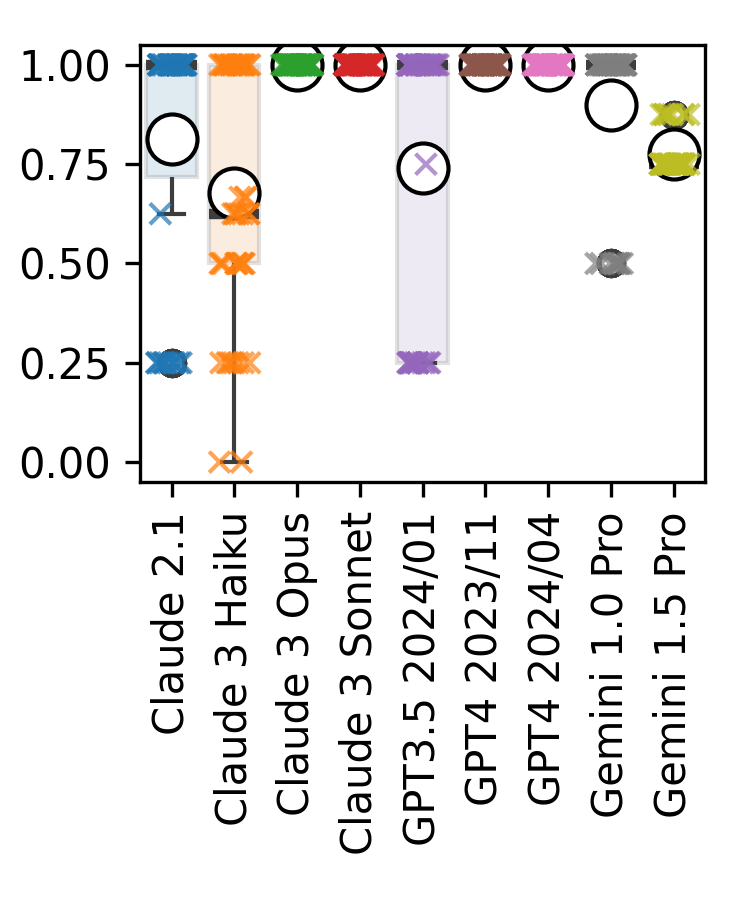}
    \caption{T2A JSON-LD}
    \label{fig:T2A-Orga-Jsonld}
  \end{subfigure}
  \hfill
  \begin{subfigure}{0.24\textwidth}
    \centering
    \includegraphics[width=\linewidth]{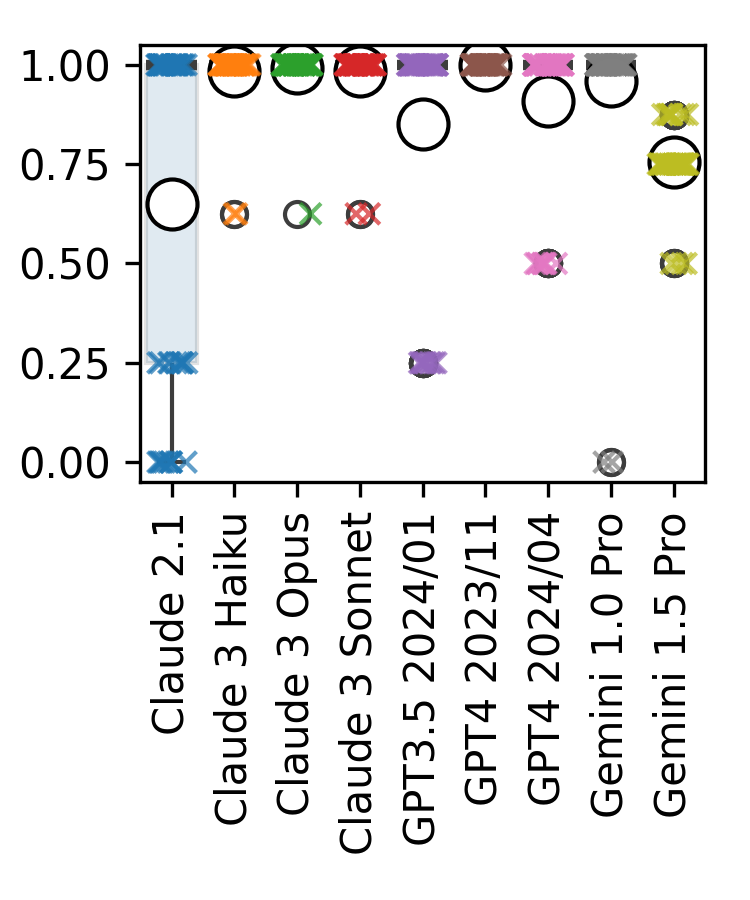}
    \caption{T2A Turtle}
    \label{fig:T2A-Orga-Turtle}
  \end{subfigure}%
  \caption{
    \emph{combinedF1} scores for the task types SPARQL to Answer (S2A) and Text to Answer (T2A) with the Organizational KG presented in JSON-LD or Turtle syntax. The score is calculated from $(f1 + trimF1 + fixedF1 + relaxedF1)/4$
  }
  \label{fig:AnswerEvalTaskScoresT2aAndS2A}
\end{figure}

\begin{figure}[h!bt]
  \begin{subfigure}{0.24\textwidth}
    \centering
    \includegraphics[width=\linewidth]{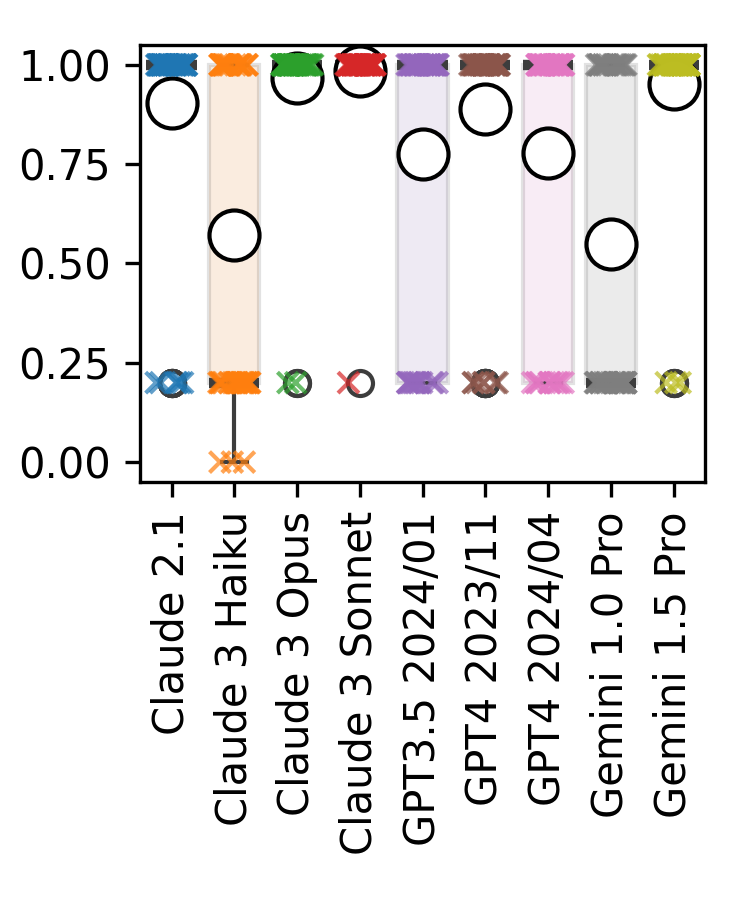}
    \caption{T2S-Orga\\(Turtle)}
    \label{fig:T2S-Orga}
  \end{subfigure}%
    \hfill
  \begin{subfigure}{0.24\textwidth}
    \centering
    \includegraphics[width=\linewidth]{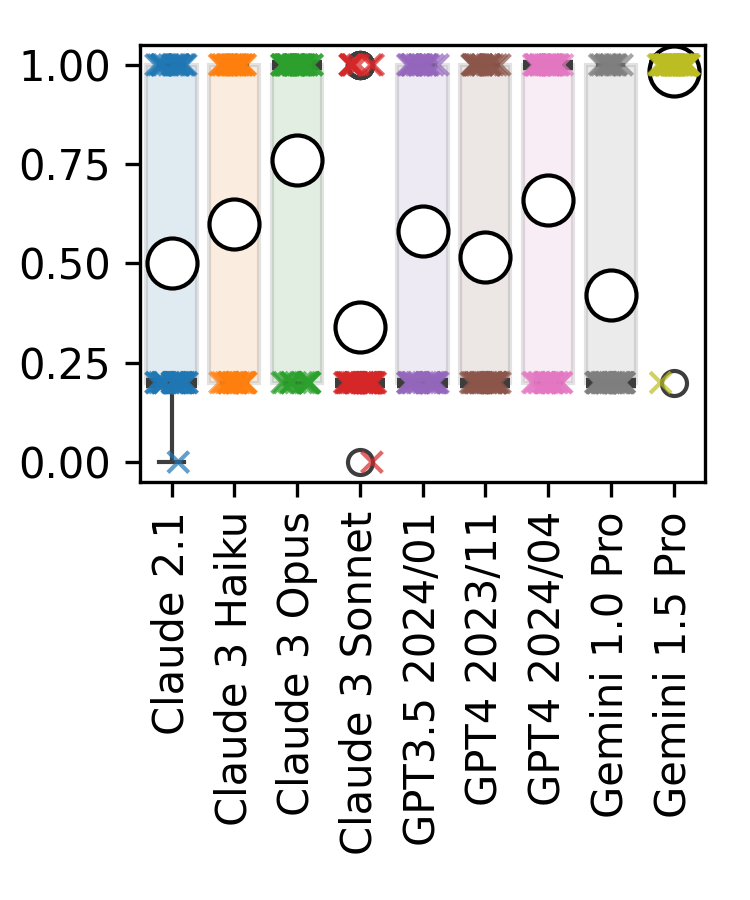}
    \caption{T2S-OrgaNum\\(Turtle+Table)}
    \label{fig:T2S-OrgaNum}
  \end{subfigure}%
  \hfill
  \begin{subfigure}{0.24\textwidth}
    \centering
    \includegraphics[width=\linewidth]{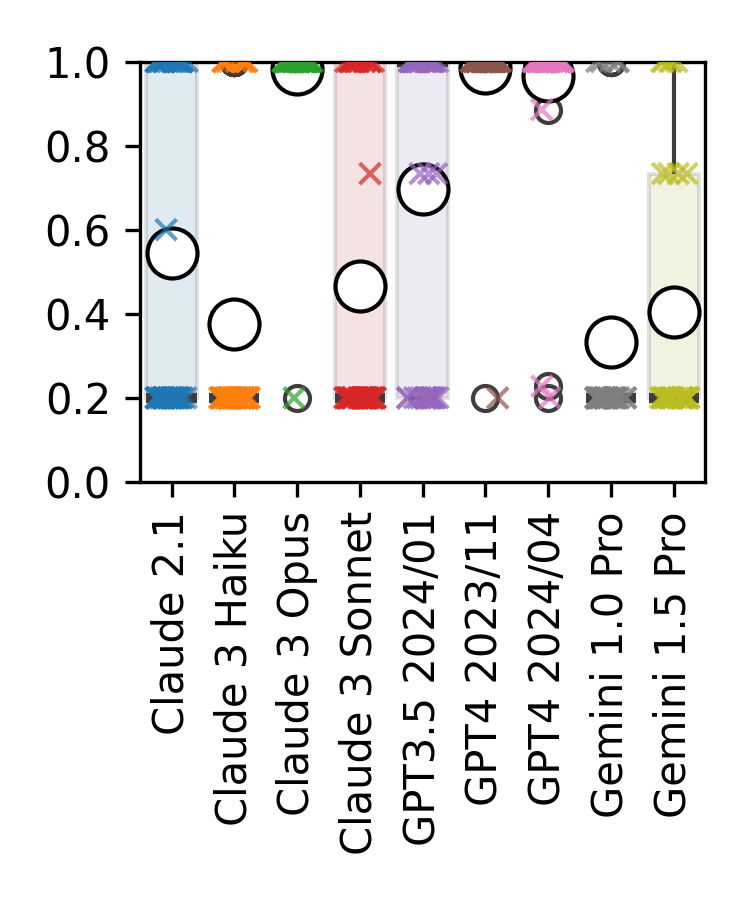}
    \caption{T2S-LC-QuAD\\(Table)}
    \label{fig:T2S-LcQuad}
  \end{subfigure}
   \begin{subfigure}{0.24\textwidth}
    \centering
    \includegraphics[width=\linewidth]{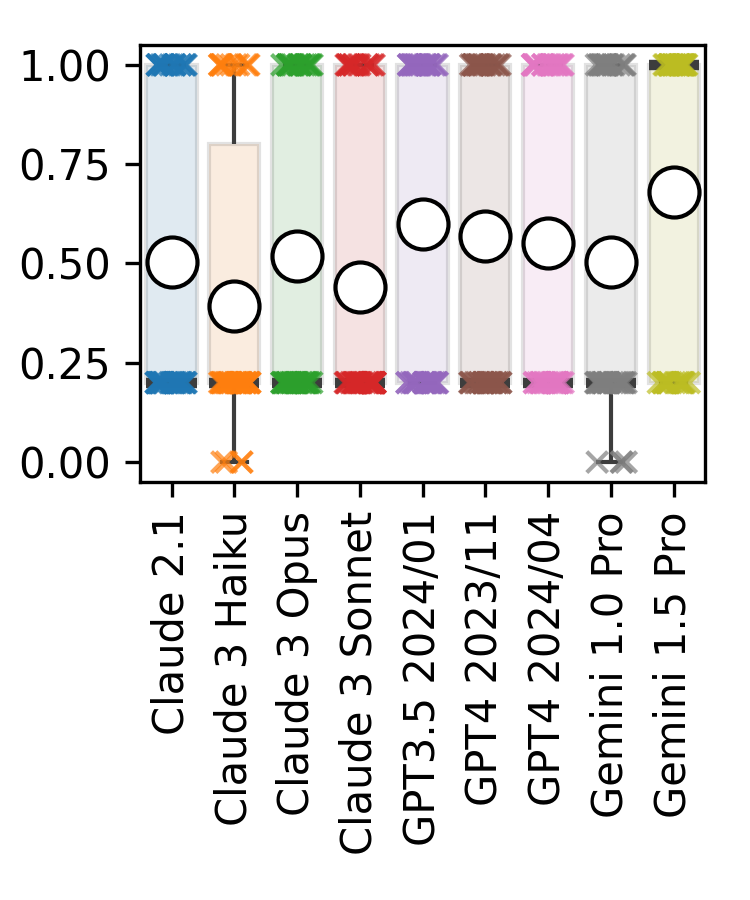}
    \caption{T2S-CoyPu\\(IRIs)}
    \label{fig:T2S-Coypu-Iris}
  \end{subfigure}
  \begin{subfigure}{0.24\textwidth}
    \centering
    \includegraphics[width=\linewidth]{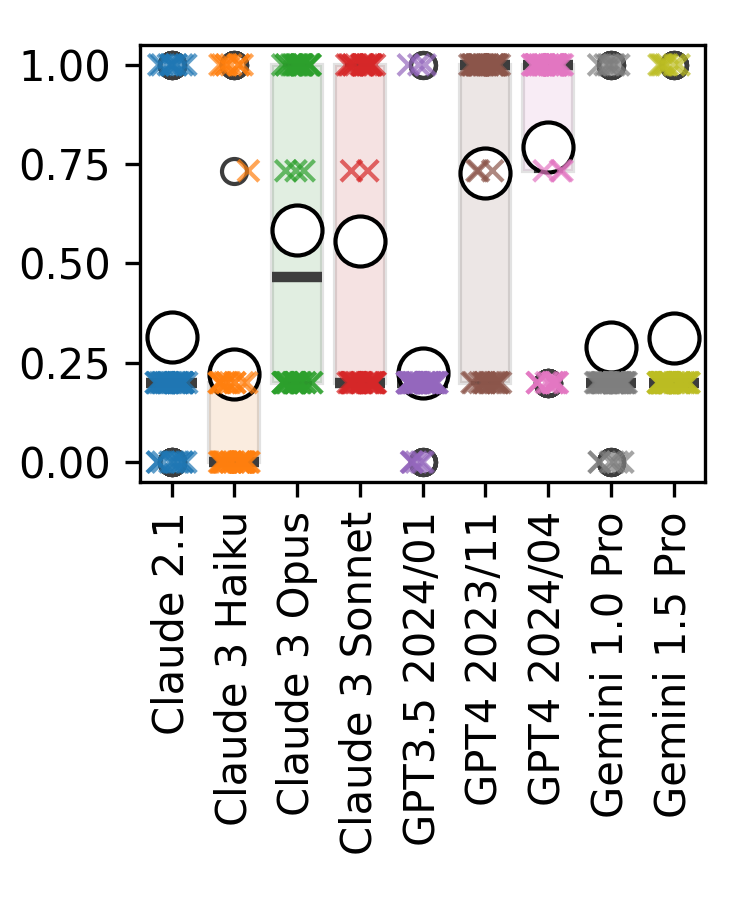}
    \caption{T2S-CoyPu\\(JSON-LD~full~KG)}
    \label{fig:T2S-Coypu-GraphJsonld}
  \end{subfigure}
  \begin{subfigure}{0.24\textwidth}
    \centering
    \includegraphics[width=\linewidth]{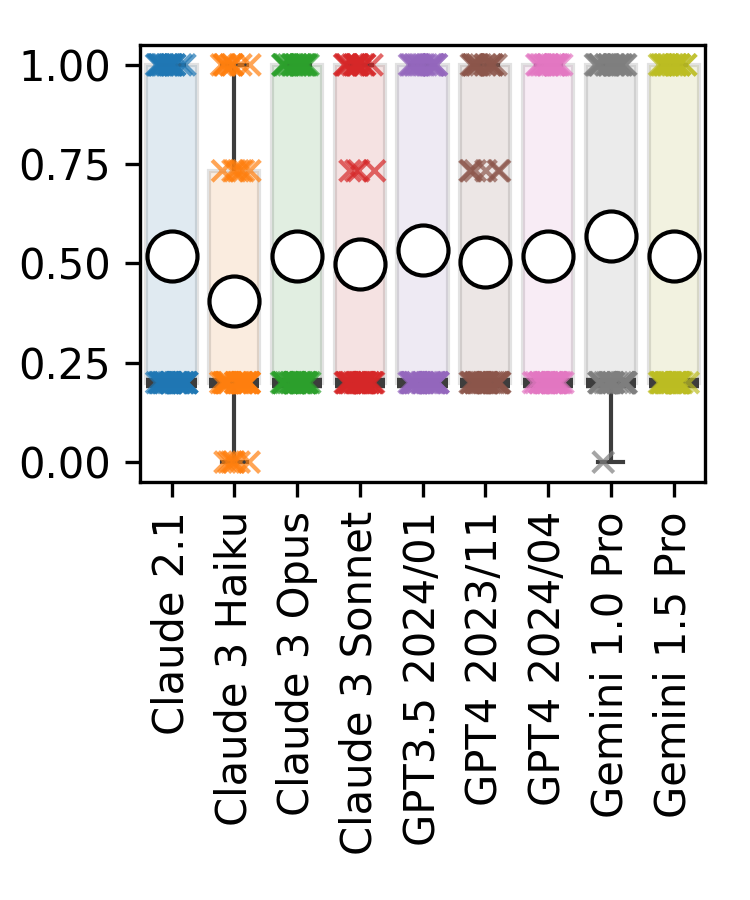}
    \caption{T2S-CoyPu\\(Turtle~full~KG)}
    \label{fig:T2S-Coypu-GraphTurle}
  \end{subfigure}
  \begin{subfigure}{0.24\textwidth}
    \centering
    \includegraphics[width=\linewidth]{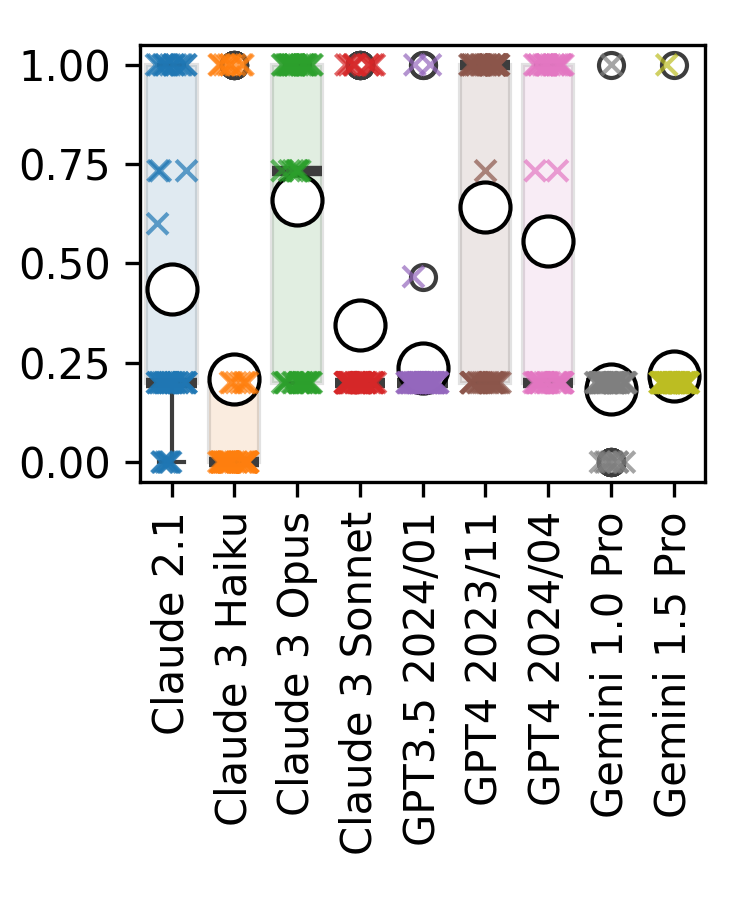}
    \caption{T2S-CoyPu\\(JSON-LD~Schema)}
    \label{fig:T2S-Coypu-SchemaJsonld}
  \end{subfigure}
  \begin{subfigure}{0.24\textwidth}
    \centering
    \includegraphics[width=\linewidth]{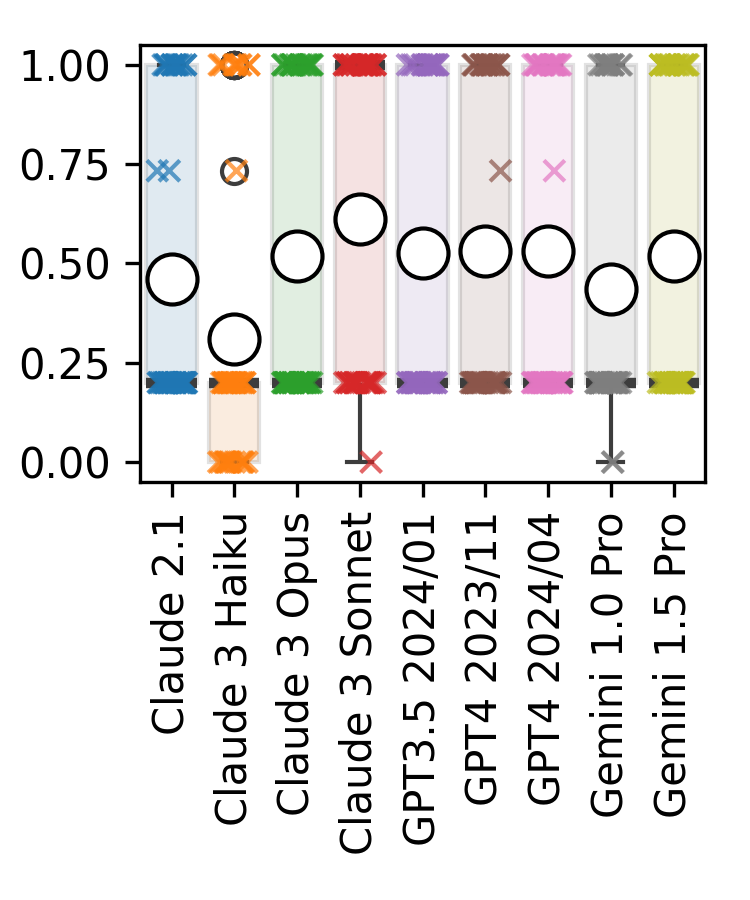}
    \caption{T2S-CoyPu\\(Turtle~Schema)}
    \label{fig:T2S-Coypu-SchemaTurtle}
  \end{subfigure}
  \begin{subfigure}{0.24\textwidth}
    \centering
    \includegraphics[width=\linewidth]{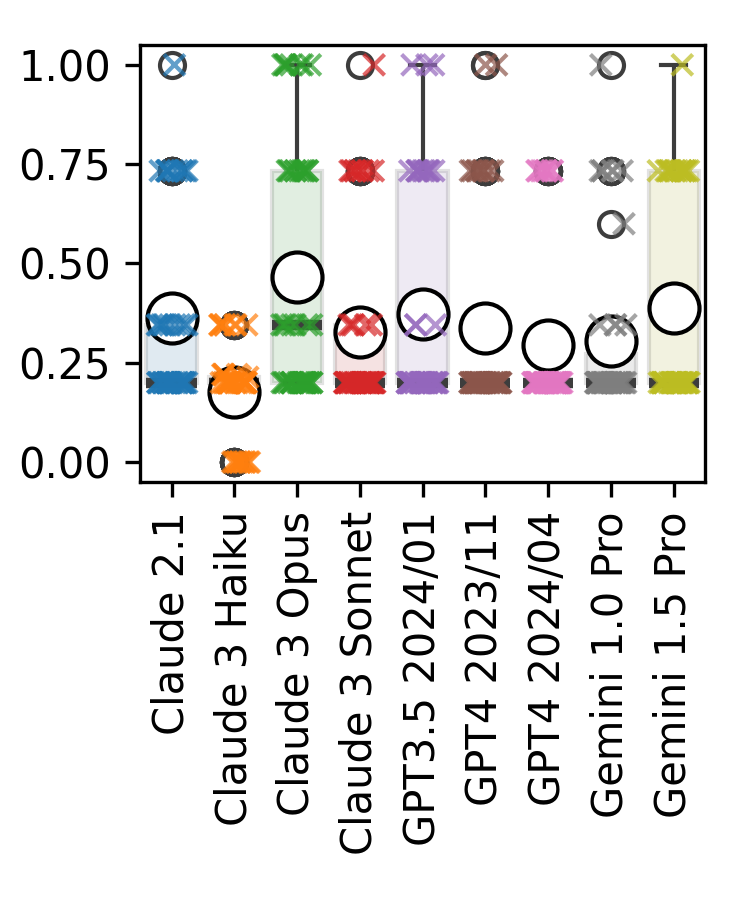}
    \caption{T2S-Beastiary\\(Turtle~Schema)}
    \label{fig:T2S-Beast-SchemaTurtle}
  \end{subfigure}
  \begin{subfigure}{0.24\textwidth}
    \centering
    \includegraphics[width=\linewidth]{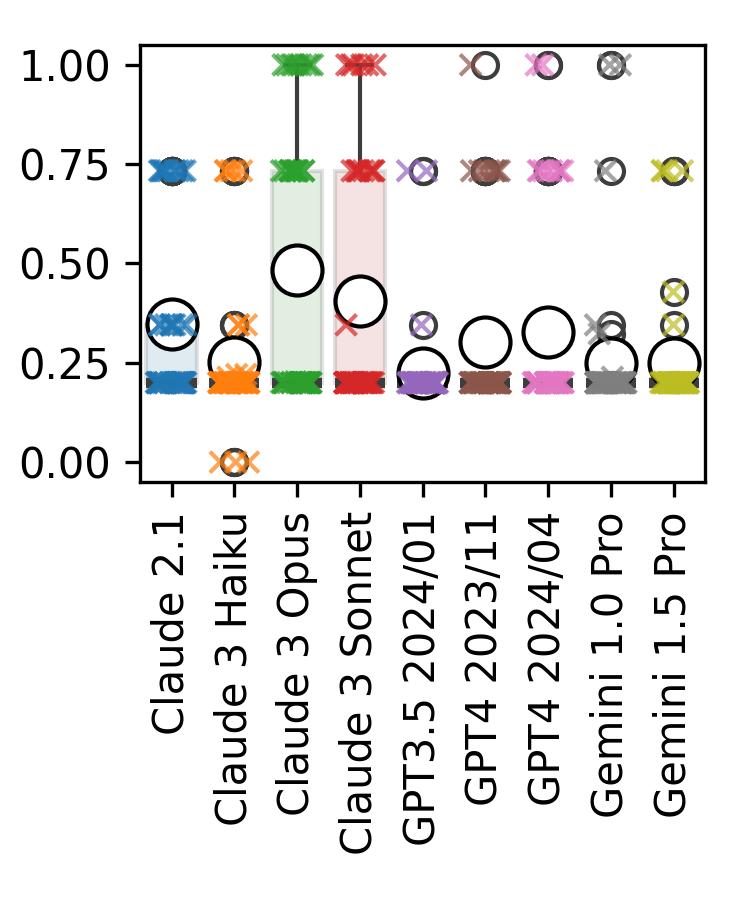}
    \caption{T2S-Beastiary\\(Turtle~Subschema)}
    \label{fig:T2S-Beast-SubschemaTurtle}
  \end{subfigure}
  \begin{subfigure}{0.24\textwidth}
    \centering
    \includegraphics[width=\linewidth]{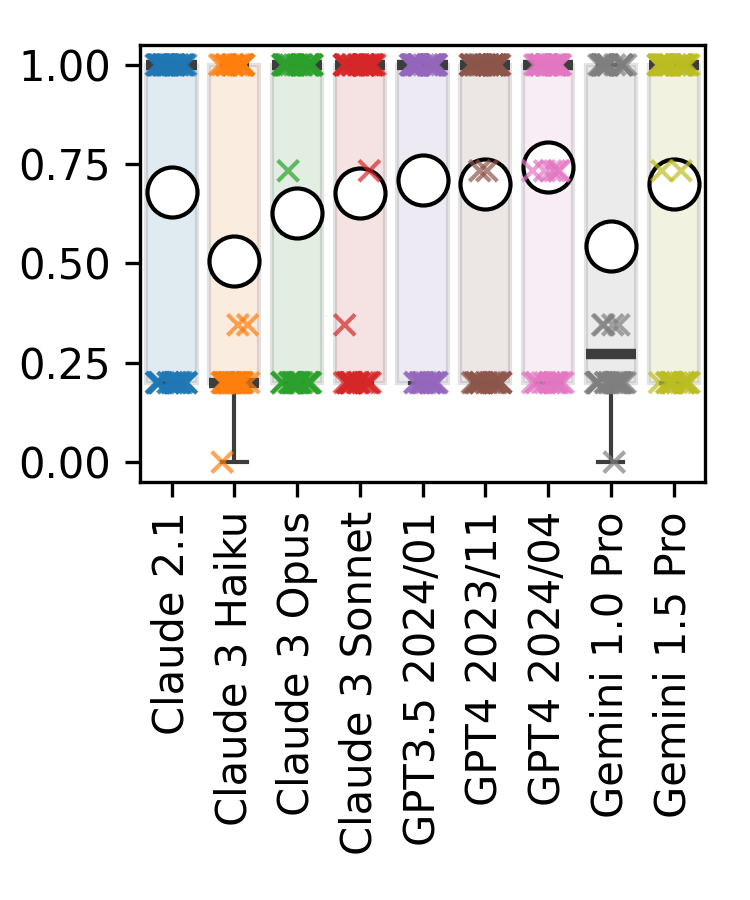}
    \caption{T2S-Beastiary\\(Turtle~KG~subset)}
    \label{fig:T2S-Beast-SubgraphTurtle}
  \end{subfigure}
    \begin{subfigure}{0.24\textwidth}
    \centering
    \includegraphics[width=\linewidth]{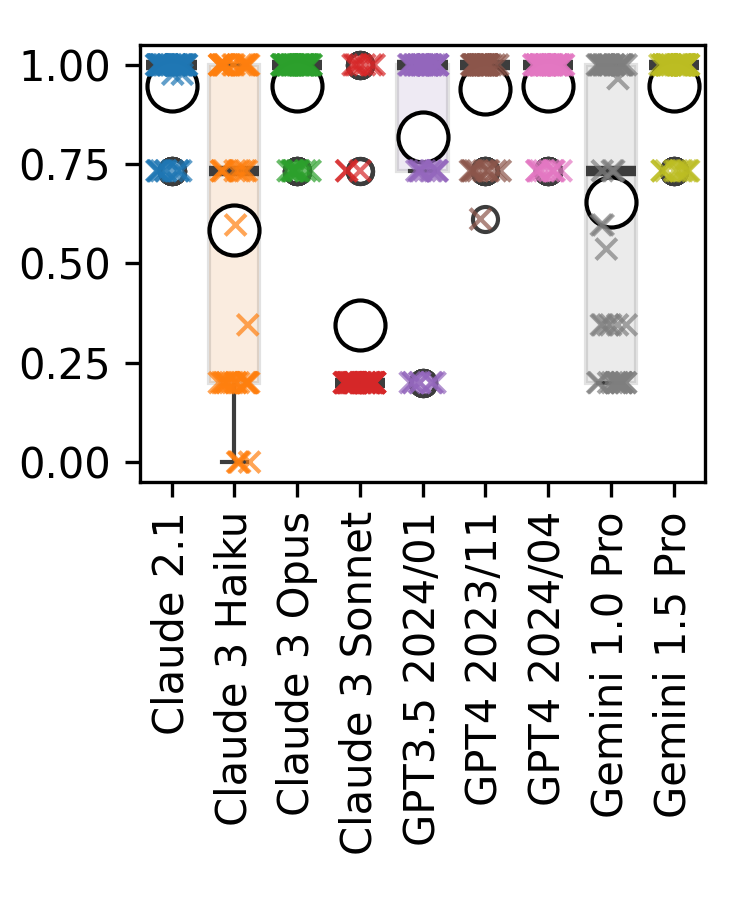}
    \caption{T2S-Beast. \newline (IRIs)}
    \label{fig:T2S-Beast-Iris}
  \end{subfigure}
  \caption{
    $max\_combined$ scores for each feedback dialog of task type Text to SPARQL (T2S).
    The combined score equals $0.2 * parse + 0.8 * f1measure$, resulting e.g. in a score of $0.2$ when f1 measure on result set is $0$ but the \sparqlselquery parses.
  }
  \label{fig:SparqlEvalTaskScores}
\end{figure}

\begin{figure}[h!bt]
  \begin{subfigure}{0.24\textwidth}
    \centering
    \includegraphics[width=\linewidth]{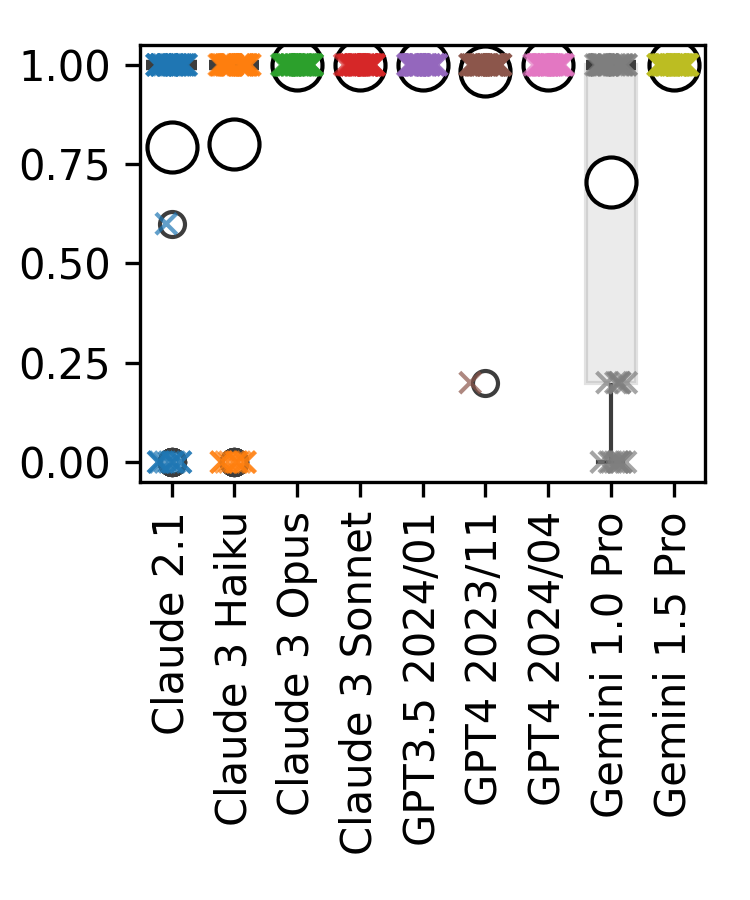}
    \caption{SSF}
    \label{fig:SSF-MaxCombined}
  \end{subfigure}
    \hfill
  \begin{subfigure}{0.24\textwidth}
    \centering
    \includegraphics[width=\linewidth]{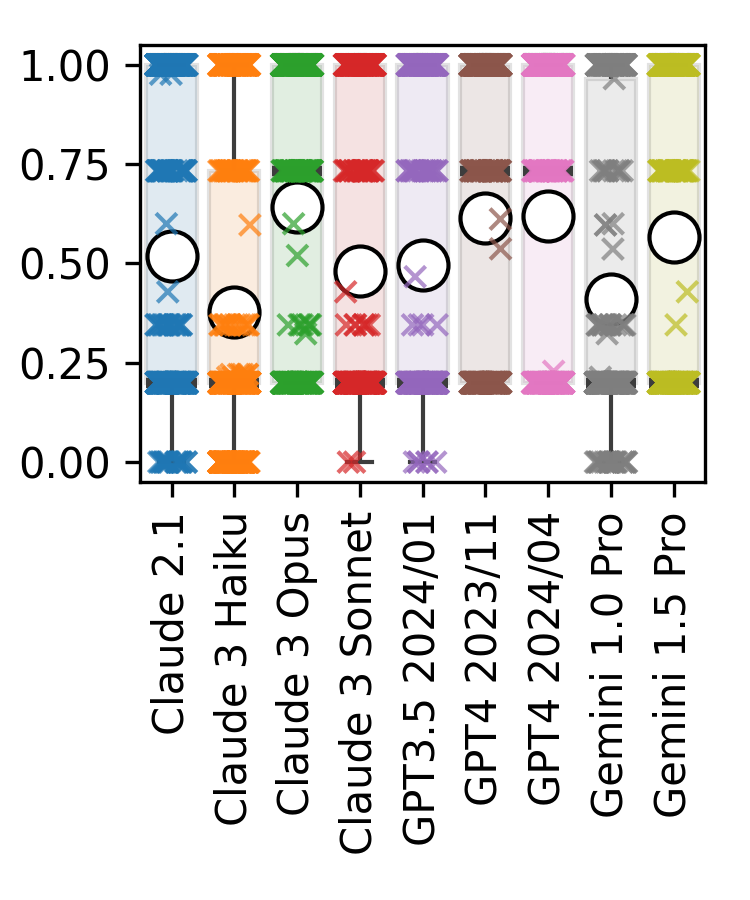}
    \caption{T2S all combined}
    \label{fig:T2S-All}
  \end{subfigure}
  \caption{
    $max\_combined$ scores for each feedback dialog of task type SPARQL Syntax Fixing (SSF) and all task type Text to SPARQL (T2S) together.
    The combined score equals $0.2 * parse + 0.8 * f1measure$, resulting e.g. in a score of $0.2$ when f1 measure on result set is $0$ but the \sparqlselquery parses.
  }
  \label{fig:SparqlSyntaxAndCombined}
\end{figure}

\subsection{Results for SPARQL Syntax Fixing (SSF)}
Fixing syntax errors in \sparqlselqueries seems to be easier for LLMs as can be seen in \cref{fig:SSF-MaxCombined}. 
Only 3 models (Claude 2.1, Claude 3 Haiku and GPT 3.5 2024/01) are not able to fix all errors within 3 feedback iterations, about 80\% of the executions got a correct answer on first try. The most common error missed was a variable name containing a hyphen. E.g. \emph{?foo-bar} was not detected as an invalid variable name.

\subsection{Results for SPARQL to Answer (S2A)}
About 75\% of the executions had perfect results and 90\% had a $relaxedF1$ score of $1$.
The most common formatting problems were caused by insertion of additional spaces.
About 10\% had just wrong answers, often for task entries that required counting.
Plots for the $combinedF1$ scores can be seen in \cref{fig:S2A-Orga-Jsonld,fig:S2A-Orga-Turtle}.

\subsection{Results for Text to SPARQL (T2S) Organizational Graph}
The state-of-the-art LLMs seem to have little problems with generating \sparqlselqueries for the organizational graph, as can be seen in \cref{fig:T2S-Orga}. 
But when we look at the results in \cref{fig:T2S-OrgaNum} for the numerical version with a translation table, we can see a different picture, where only Gemini 1.5 scores almost perfect.

\subsection{Results for Text to SPARQL (T2S) LC-QuAD}
In \cref{fig:T2S-LcQuad} we can see that GPT4 in both versions and Claude 3 Opus were quite successful in generating correct \sparqlselqueries. 
They managed to generate syntactically correct queries in 99$\%$ of the cases on the first try, with all queries having correct syntax after the final feedback iteration.
The $f1measure$ score for these models increased from $0.699$ for their first answer to $0.972$ which means they really benefited from the provided feedback and knowledge graph information.
Gemini 1.0 on the other hand reached only an $f1$ score of $0.274$ after the final feedback loop, indicating that it still struggles to use the provided properties correctly.
However the queries were syntactically correct in almost all cases.
For the other models the property structure of Wikidata seems to be just as challenging.
We often see generated graph patterns with properties used in a reversed direction (domain and range swapped).

\subsection{Results for Text to SPARQL (T2S) CoyPu}

For the CoyPu-Mini graph, the LLMs struggled more to generate syntactically correct SPARQL queries in the \textbf{first answer}.
When being given the IRIs, Haiku alone is responsible for 21 of the 45 errors, with Gemini generating 8 syntactically wrong queries and GPT4 just one (2023/11 model).
The rest is shared between other flavors of Claude.
The situation becomes worse when we provide the full KG; only 316 of the responses for JSON-LD and 401 for Turtle contained valid SPARQL (with 450 in total for each).
We get the same picture if we give the Schema information, with 341 valid SPARQL queries for JSON-LD and 397 for Turtle.
However, if we provide feedback and allow LLMs to correct their mistakes, the numbers increase to 391 for the JSON-LD graph and 434 for Turtle.
All model families (GPT, Gemini, Claude) have instances where even the last answer they gave contained no valid SPARQL, so this is a phenomenon that can happen with each LLM.
Looking at the $last_f1measure$ score (the final answer) of each LLM 
we found that the global average is about $0.3$ for JSON-LD schema information
and $0.4$ for every other kind of information passed.
This is valid across all models, except for one exception: The $last_f1measure$ score of the GPT4 models increases to $0.7$ if we give it the full KG in JSON-LD format.
Surprisingly this did not happen with the Turtle format.
The plots for the $max\_combined$ in \cref{fig:T2S-Coypu-GraphJsonld,fig:T2S-Coypu-GraphTurle,fig:T2S-Coypu-Iris,fig:T2S-Coypu-SchemaJsonld,fig:T2S-Coypu-SchemaTurtle}.
show that the Gemini models, Claude 2.1 and Haiku perform worse for JSON-LD compared to using Turtle representations of the full KG or schema opposed to Opus and the GPT4 versions that perform better with JSON-LD.

\subsection{Results for Text to SPARQL (T2S) Beastiary}

Generating syntactically correct SPARQL queries was no problem for the LLMs (besides Haiku) as can be seen in \cref{fig:T2S-Beast-SchemaTurtle,fig:T2S-Beast-SubschemaTurtle,fig:T2S-Beast-SubgraphTurtle,fig:T2S-Beast-Iris}.
No matter what kind of information was given about the KG, there were always about 420 of 450 queries that were parseable.
Again, GPT4 performs best here with only answering two times with syntactically wrong queries (at the end of the dialog), and Haiku being responsible for about half of the errors, the rest being evenly distributed among the other flavors of Claude as well as Gemini.
Looking at the $f1$ score we can see that the kind of information provided has a large influence on the correctness of the query.
Providing the schema in Turtle format yields a global $max\_f1measure$ of $0.19$ across all models, with the sub schema scoring only a $0.16$.
Providing a relevant portion of the KG as a sub graph however yields a score of $0.61$ and providing the IRIs needed for the query nets a $0.76$.

\subsection{Results for Text to Answer (T2A)}
About 75\% of the answers given were perfectly correct and about 97\% good enough for the relaxed scoring.
The most common formatting problem were the addition of spaces in the given answer.
The few really wrong answers given were mainly on a question asking for a count, and the LLM answered with the wrong number.
Plots for the $combinedF1$ score are given in \cref{fig:T2A-Orga-Jsonld,fig:T2A-Orga-Turtle}.

\subsection{Statistical Analysis}

The above sections deal with the discussion of the results on a split-by-task basis, but we can also analyze the results based on different aspects that were covered, like the input format of a knowledge graph (\textit{JSON-LD} vs \textit{Turtle}) and the kind of information that was provided to the LLM (Full KG vs Schema information vs List of IRIs).
We ran pairwise t-Tests and obtained the results shown in \cref{tab:t_stats}.
For each comparison we took the f1 scores (\verb|mean_f1measure|) of two aspects across all tasks and performed a Welch's t-Test with the two resulting populations, labeled $A$ and $B$.
In each case the null hypothesis $H_0$ was that $\mu_A = \mu_B$.
Using the common threshold of $p=0.05$ we can see that the null hypothesis can be rejected every time except for the difference between providing a list of relevant IRIs to the LLM vs the full KG.
In the three remaining cases, we can see that there is a statistically significant difference in the data: Turtle vs JSON-LD, IRIs vs Schema, full KG vs Schema.

\begin{table}[]
    \caption{The $t$ statistic and corresponding $p$ value for two populations, obtained by grouping the data from all tasks on two aspects ($A$ and $B$) each and taking the $f1$ score from \textit{mean\_f1measure}.}
    \centering
    \begin{tabular}{c c  c  c}
    \toprule
         Aspect $A$ & Aspect $B$ & $t$ & $p$ \\ \hline
         Turtle  & JSON-LD & $3.87$ & $0.00011$ \\
         IRIs  & Schema & $3.00$ & $0.0027$ \\
         IRIs  & full KG & $1.14$ & $0.254$ \\
         Graph  & Schema & $2.29$ & $0.022$ \\
    \bottomrule
    \end{tabular}

    \label{tab:t_stats}
\end{table}

\section{Conclusion and Outlook}
\label{sec:conclusion}

Benchmarking LLMs on \sparqlselquery related tasks is difficult and more experiments are needed.
Nonetheless with the work presented in this paper we contributed to the research questions defined and layed the ground for further automated evaluation.

From the results gathered we can give the following answers to the research questions:
Most LLMs evaluated had no problems with working with \sparqlselquery syntax (RQ1) or reading \sparqlselquery semantics (RQ2).
Creating \sparqlselqueries with correct semantics (RQ3) seems to be still a difficult task for the LLMs evaluated here, at least when applying the approach to evaluate the answer generated.
The results seem to depend on different aspects (RQ4).
The varying performance between the 10 different variants of T2S tasks indicate, that we should further extend the tests for serialization formats and content that is provided as KG information to the LLMs (full graph vs. sub graph vs. schema vs. IRIs).
Taking the CoyPu-Mini KG as an example, we found that Turtle leads to a higher number of syntactically correct queries, whereas JSON-LD improved the score of GPT4 tremendously and made it stand above every other model in this specific test.

The results of LLMs depend a lot on the concrete task setting as can be seen in \cref{fig:SparqlEvalTaskScores}.
Thus we found no clear \emph{winner}.
But when focusing on the mean value of f1 scores for T2S tasks Claude 3 Opus and GPT 4 scored best as shown in \cref{fig:T2S-All}.

Further research could improve the evaluation of \sparqlselqueries created by LLMs, especially for queries returning no result when applied on the KG.

In the long term, the fine-tuning or training of LLMs would probably benefit from more SPARQL-related training data.

Unfortunately, we found that existing KGQA benchmarks have ambiguity that hinder a proper automatic evaluation.
Given the fact those publicly available benchmarks (like LC-QuAD) can be contained in the training dataset and memorized by the LLMs, emphasizes the need for new and diverse test datasets.
Fortunately, with the extended framework presented in this work, it is easy to include new evaluation datasets.

\section*{Acknowledgments}
This work was partially supported by grants from the German Federal Ministry of Education and Research (BMBF) to the project StahlDigital (13XP5116B) and ScaleTrust (16DTM312D) as well as from the German Federal Ministry for Economic Affairs and Climate Action (BMWK) to the KISS project (01MK22001A) and CoyPu project (01MK21007A).

\section*{Conflicts of Interest}
The authors have no competing interests to declare that are relevant to the content of this article.
The authors used a free evaluation license for Claude and Gemini models, however due to the setup and technical nature of the evaluation this has no effect on the results.

\section*{Online Resources}
\begin{description}
    \item[code:] \url{ https://github.com/AKSW/LLM-KG-Bench/tree/v2.0.0}, \href{https://doi.org/10.5281/zenodo.13622575}{DOI:10.5281/zenodo.13622575}
    \item[results:] \url{https://github.com/AKSW/LLM-KG-Bench-Results/tree/main/2024-NLP4KGC-SPARQL}, \href{https://doi.org/10.5281/zenodo.13621581}{DOI:10.5281/zenodo.13621581}
\end{description}

\bibliography{refs.bib}

\clearpage
\pagestyle{plain}
\cfoot*{} 

\section*{Metadata for this article}

\begin{description}
    \item[Title:] Assessing SPARQL capabilities of Large Language Models
    \item[Authors:] Lars-Peter Meyer, Johannes Frey, Felix Brei and Natanael Arndt
    \item[workshop:] \href{https://sites.google.com/view/3rdnlp4kgc}{Natural Language Processing for Knowledge Graph Creation 2024 (NLP4KGc) @~SEMANTiCS~2024}, 17.-19.~9.~2024 in Amsterdam, Netherlands
    \item[original publication:] \url{https://ceur-ws.org/Vol-3874/paper3.pdf}
    \item[original publication date:] 19. 12. 2024
    \item[submitted for review:] 26.~7.~2024
    \item[peer review status:] accepted by peer review (23.~8.~2024)
    \item[submitted for publication:] 31.~8.~2024
    \item[code:] \url{ https://github.com/AKSW/LLM-KG-Bench/tree/v2.0.0}, \href{https://doi.org/10.5281/zenodo.13622575}{DOI:10.5281/zenodo.13622575}
    \item[results:] \url{https://github.com/AKSW/LLM-KG-Bench-Results/tree/main/2024-NLP4KGC-SPARQL}, \href{https://doi.org/10.5281/zenodo.13621581}{DOI:10.5281/zenodo.13621581}
    \item[Bibtex entry:] 
\end{description}

\begin{scriptsize}
  \begin{verbatim}
@inproceedings{Meyer2024AssessingSparqlCapabilititesLLM,
  author = {Meyer, Lars-Peter and Frey, Johannes and Brei, Felix and Arndt, Natanael},
  title = {Assessing {SPARQL} capabilities of Large Language Models},
  year = {2024},
  booktitle = {Proceedings of the 3rd International Workshop on Natural Language Processing for
    Knowledge Graph Creation co-located with 20th International Conference on Semantic Systems
    ({SEMANTiCS} 2024)},
  editor = {Vakaj, Edlira and Iranmanesh, Sima and Stamartina, Rizou and 
      Mihindukulasooriya, Nandana and Tiwari, Sanju and Ortiz-Rodríguez, Fernando and Mcgranaghan, Ryan},
  pages = {35–-53},
  series = {{CEUR} Workshop Proceedings},
  volume = 3874,
}    
  \end{verbatim}
\end{scriptsize}

\end{document}